\documentclass[preprint]{aastex}
\pdfoutput=1
\usepackage{graphics,color}
\usepackage[utf8]{inputenc}
\usepackage{wrapfig}

\begin{document}

\title{Depletions of Elements from the Gas Phase: A Guide on Dust 
Compositions}

\author{Edward B. Jenkins}
\affil{Princeton University Observatory\\
Princeton, NJ 08544-1001}
\email{ebj@astro.princeton.edu}

\begin{abstract}
Ultraviolet spectra of stars recorded by orbiting observatories since the 1970’s have revealed 
absorption features produced by atoms in their favored ionization stages in the neutral ISM of our 
Galaxy.   Most elements show abundances relative to hydrogen that are below their values in stars, 
indicating their removal by condensation into solid form.  The relative amounts of these 
depletions vary from one location to the next, and different elements show varying degrees of 
depletion.  In a study of abundances along 243 different sight lines reported in more than 100 
papers, Jenkins (2009) characterized the systematic patterns for the depletions of 17 different 
elements, and these results in turn were used to help us understand the compositions of dust 
grains.  Since the conclusions are based on differential depletions along different sightlines, they 
are insensitive to errors in the adopted values for the total element abundances.  Some of the more 
remarkable conclusions to emerge from this study are that (1) oxygen depletions in the denser gas 
regions (but not as dense as the interiors of molecular clouds) are stronger than what we can 
expect from just the formation of silicates and metallic oxides, and (2) the chemically inert 
element krypton shows some evidence for weak depletion, perhaps as a result of trapping within 
water clathrates or binding with ${\rm H}_3^+$.
\end{abstract}

\section{Background}\label{background}

\begin{wrapfigure}{r}{0.5\textwidth}
\begin{center}
\includegraphics[width=0.48\textwidth]{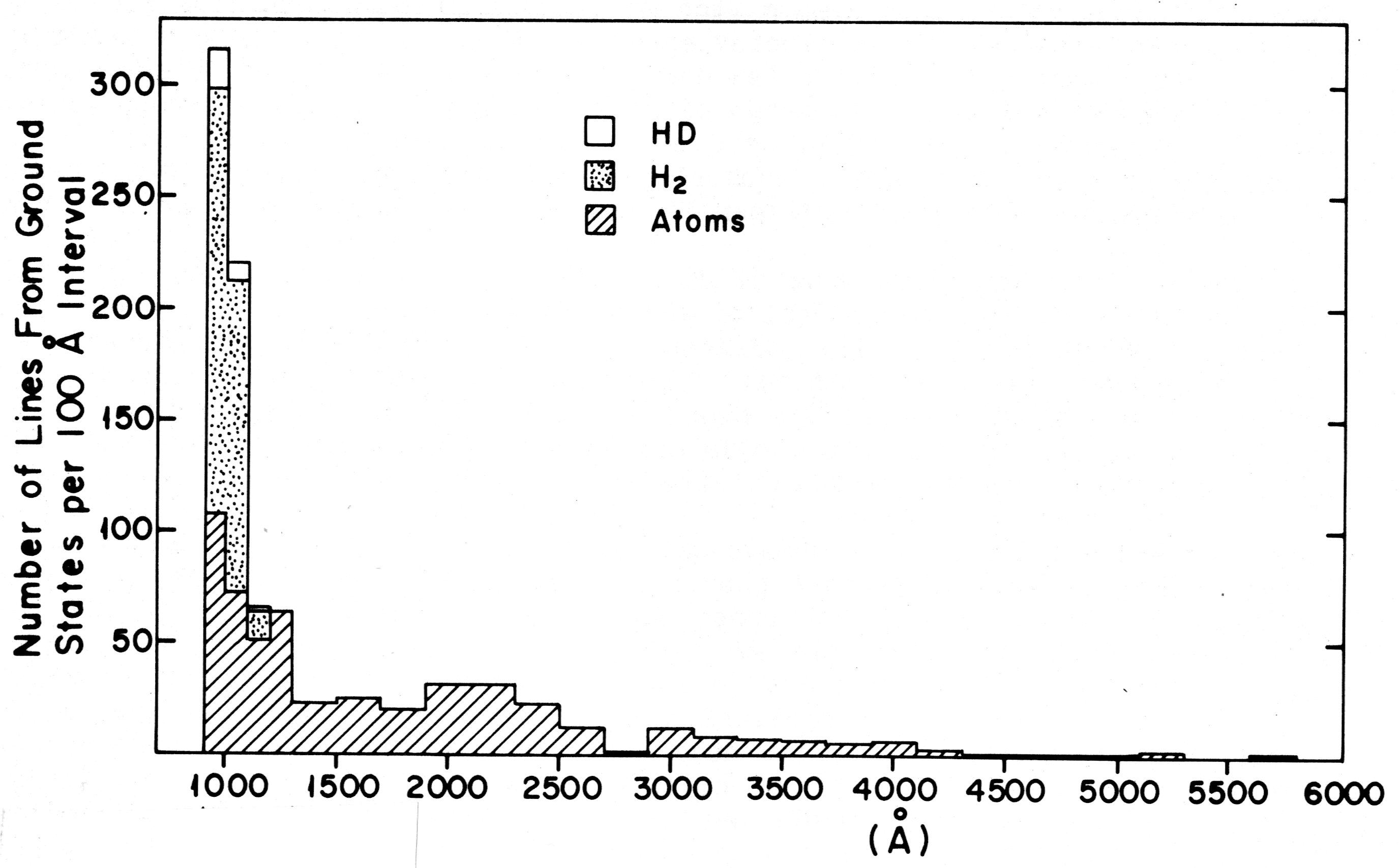}
\end{center}
\caption{Line density vs. wavelength\label{line_density}}
\end{wrapfigure}

When various elements condense into solid form in the interstellar medium, there 
are profound reductions in the relative abundances of many types of heavy 
atoms and ions in the gas phase. We can obtain insights on the compositions of 
the dust grains by studying the strengths of depletions of different gas 
constituents below what we believe to be their intrinsic abundances.  Virtually all 
of the free atoms in the neutral interstellar medium are in their ground electronic 
states.  For nearly all of the important atomic species, the energy separations of 
excited levels above the ground state are larger than the energies of transitions 
in the visible part of the spectrum.  Hence, as shown in Fig.~\ref{line_density}, for 
absorption spectroscopy a vast majority of the useful features are situated in the 
UV part of the spectrum; this fact was recognized long ago by Spitzer and 
Zabriskie (1959).  For studying gas in our Galaxy or nearby systems, these 
transitions can only be accessed by observatories in space.  Except for the 
elements Be (Spitzer 1949 ; Herbig 1968 ; Boesgaard 1985) and Ti (Wallerstein 
\& Goldsmith 1974 ; Stokes \& Hobbs 1976 ; Stokes 1978 ; Wallerstein \& Gilroy 
1992 ; Welsh et al. 1997 ; Prochaska et al. 2005 ; Ellison et al. 2007 ; Welty \& 
Crowther 2010) and some comparisons of the abundances of Ca to Na with 
velocity (Routly \& Spitzer 1952 ; Siluk \& Silk 1974 ; Vallerga et al. 1993 ; 
Sembach \& Danks 1994), research on interstellar atomic abundances relied 
almost entirely on recording stellar spectra over the range extending from the 
Lyman limit (912$\,$\AA) up to about 3000$\,$\AA, where the Earth's atmosphere 
becomes transparent. 

The {\it Copernicus\/} satellite (Rogerson et al. 1973), launched in 1972, was the 
first major observatory that had the resolution and sensitivity for productive 
research on atomic abundances in the interstellar medium (Spitzer \& Jenkins 
1975 ; Cowie \& Songaila 1986).  The earliest results from this facility (Morton et 
al. 1973) exhibited a pattern of element depletions that led Field (1974) to 
conclude that the severities of depletions are strongly linked to the temperatures 
at which the respective elements condense into solid compounds in matter being 
expelled from the outer portions of stellar atmospheres.  This formation 
mechanism, along with the creation of dust in supernova ejecta and the further 
growth of grains in dense gas complexes in space, still remains a central theme 
in theories of dust production.  A decade of observations with {\it Copernicus\/} 
reached a climax with the study of element abundances across a broad range of 
sight lines toward bright stars in our Galaxy (Jenkins et al. 1986), revealing that 
the strengths of depletions were strongly linked to the average gas densities, 
indicating that the growth and destruction of grains depend on local conditions.

Following the era of {\it Copernicus\/}, major advances in achieving higher 
accuracy and viewing fainter stars arose from observations by the {\it Goddard 
High Resolution Spectrograph\/} (Brandt et al. 1994) aboard the {\it Hubble 
Space Telescope} (HST) soon after its launch in 1990 (Cardelli et al. 1991a ; 
Cardelli et al. 1991b ; Savage et al. 1991 ; Savage et al. 1992 ; Cardelli et al. 
1993 ; Sofia et al. 1993 ; Cardelli 1994).  Savage and Sembach (1996) provided 
an excellent review of these early accomplishments with HST.  A major upgrade 
in the ability to record a broad wavelength coverage at high resolution was 
provided when the {\it Space Telescope Imaging Spectrograph\/} was installed on 
the second servicing mission of HST in 1997 (Woodgate \& et al. 1998).  The {\it 
Far Ultraviolet Spectroscopic Explorer\/} (FUSE), launched in 1999, extended our 
coverage to a wavelength band between the Lyman limit and the short-
wavelength cutoff of HST.  This facility allowed us to determine the amount of 
H$_2$ that accompanied the H~I gas and also gave us an access to some 
important atomic transitions that were not available to HST (Moos et al. 2000).

\section{A Unified Representation of Depletions}\label{unified}
\begin{wrapfigure}{r}{0.4\textwidth}
\vspace{-1.cm}
\begin{center}
\includegraphics[width=0.38\textwidth]{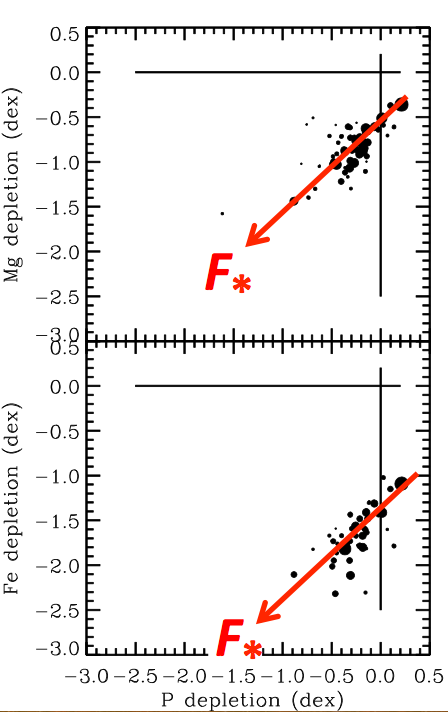}
\end{center}
\vspace{-.5cm}
\caption{Interstellar atomic depletion factors [Mg$_{\rm gas}$/H] ({\it top panel\/}) 
and [Fe$_{\rm gas}$/H] ({\it bottom panel\/}) vs. [P$_{\rm gas}$/H].  
Measurements with small errors are depicted with disks that have large 
diameters, while those with large errors are more point-like.  The red trend line 
labeled $F_*$ shows the progression from mild depletions to more severe ones 
(for all elements), depending on how far grain development has progressed for a 
particular sight line.\label{depletion_comparison}}
\end{wrapfigure}

During the first 35 years of observations using the orbiting observatories to 
gather absorption line data, there had emerged a large number of publications 
that reported on a heterogeneous mix of different element abundances over a 
diverse collection of sight lines.  The results by themselves were of significant 
value and led to interesting conclusions.  Nevertheless, there still remained a 
need to consolidate the information from different sources and then 
systematically describe the outcomes in terms that linked the depletion of each 
element to those of the others, and at the same time, accounted for the variations 
that were known to take place from one region to the next. This task was 
accomplished in 2009 by Jenkins (2009), who interpreted the abundances of 17 
different elements reported in more than 100 papers that covered 243 sight lines 
to different stars.  He developed a simple unification scheme that described in 
straightforward parametric forms the logarithmic depletion factor of any element 
$X$ from the gas phase,
\begin{equation}\label{depl_def}
[X_{\rm gas}/{\rm H}]=\log\left({N(X)\over N({\rm H})}\right)_{\rm obs} - 
\log\left({X\over {\rm H}}\right)_{\rm ref}~,
\end{equation}
where $N(X)$ is the column density of element $X$, $N({\rm H})$ represents the 
column density of hydrogen in both atomic and molecular form, i.e., $N({\rm 
H~I})+2N({\rm H}_2)$, and the reference abundance ratio $(X/{\rm H})_{\rm ref}$ 
can be taken from either the abundances measured for B-type stars or the Sun.  
It follows that the missing atoms that make up the relative contribution of element 
$X$ in the form of dust grains or large molecules is given (in linear terms) by the 
expression
\begin{equation}\label{x_dust}
(X_{\rm dust}/{\rm H})=(X/{\rm H})_{\rm ref} (1-10^{[X_{\rm gas}/{\rm H}]})~.
\end{equation}
Aside from constructing a unification scheme for element depletions that could 
lead to dust compositions, Jenkins had to exercise many quality control 
procedures and devise corrections to the earlier results that accounted for 
updated values for the transition strengths.

The basic foundation for the construction of a unified representation of depletions 
is based on the observation that any particular combination of two elements 
exhibits logarithmic depletions that have a linear relationship with each other.  
This concept is illustrated in Fig.~\ref{depletion_comparison}.  When compared to 
the logarithms of the depletions of P, the logarithmic depletions of Mg and Fe 
show a linear trend.  Different sight lines have varying degrees of depletions of all 
three elements, and the overall severity of such depletions for any sight line can 
be characterized by a single parameter, which Jenkins designated as $F_*$.  
The differences in the slopes and intercepts of the trend lines representing a 
progression of $F_*$ can be represented by just two additional parameters that 
are unique to each element.  The $F_*$ sight-line parameter depends only on 
observed depletions, and it replaces the average density $\langle n({\rm 
H})\rangle\equiv [N({\rm H~I})+2N({\rm H}_2)]/d$ or molecular hydrogen fraction 
$f({\rm H}_2)\equiv 2N({\rm H}_2)/[N({\rm H~I})+2N({\rm H}_2)]$ as a 
discriminant that had been used by earlier investigators (Jenkins, et al. 1986 ; 
Cardelli 1994 ; Snow et al. 2002).  The adopted scale for $F_*$ was arbitrary: 
Jenkins defined $F_*=0.0$ to correspond to the weakest depletions that 
appeared in the survey, while $F_*=1.0$ represented the depletions seen toward 
the well studied, strongly depleted $-15\,{\rm km~s}^{-1}$ component toward the 
star $\zeta$~Oph (Savage, et al. 1992).  Occasionally, sight lines exhibited 
values of $F_*$ outside the range $0.0-1.0$.

\begin{figure}[h]
\begin{center}
\includegraphics[width=5in]{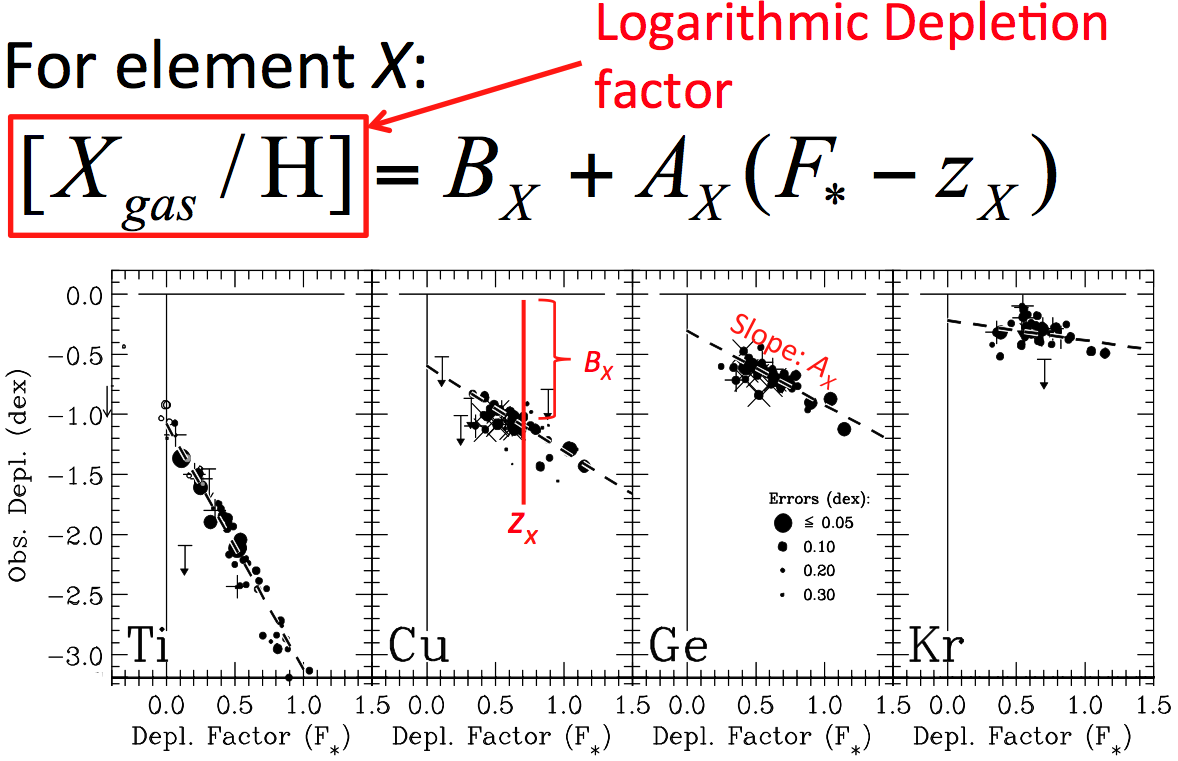}
\caption{Examples for 4 different elements that illustrate the linear equation that 
relates the logarithmic depletion factor for an element $X$ to the sight-line 
parameter $F_*$.  The zero point $z_X$ for $F_*$ is chosen such that, with the 
observations and their errors at hand, the covariance of the errors in the other 
two parameters, $A_X$ (slope) and $B_X$ (vertical offset) is 
zero.\label{basic_equation}}
\end{center}
\end{figure}
\begin{wrapfigure}{r}{0.4\textwidth}
\vspace{-2.5cm}
\begin{center}
\setlength\fboxsep{0pt}
\setlength\fboxrule{1.5pt}
\fbox{\includegraphics[width=0.38\textwidth]{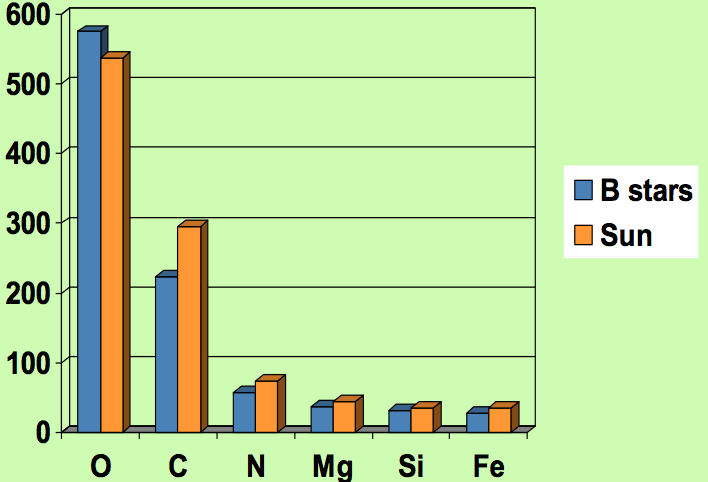}}
\end{center}
\vspace{-.2cm}
\caption{A comparison of B-star (Nieva \& Przybilla 2012) and solar (Asplund, et 
al. 2009) reference abundances for some selected elements.  The units on the 
$y$-axis correspond to parts per million H atoms.\label{ref_abundances}}
\end{wrapfigure}

Figure~\ref{basic_equation} illustrates the manner by which Jenkins 
parameterized the depletion trends for different elements. Table~\ref{parameters} 
lists the values of $A_X$, $B_X$, and $z_X$ for the 17 elements in the study, 
along with the adopted reference abundances (taken to be solar abundances 
specified by Lodders (2003)).  One drawback of having to rely on reference 
abundances is the fact that some of them are controversial.  
Figure~\ref{ref_abundances} shows that there are some disagreements between 
B-star abundances (Nieva \& Przybilla 2012) and solar abundances (Asplund et 
al. 2009).  [We can ask the question ``Why is it that for many heavy elements the 
abundances in the Sun (4.5~Gyr old) appear to higher than those in young B 
stars (several tens of Myr old)?''  The answer may arise from a suggestion by 
Lyubimkov (2013) that ionizations in the standard stellar atmosphere models for 
early-type stars are underestimated.  From Eq.~\ref{x_dust} it should be clear 
that the answers that we will obtain for the dust elemental compositions will 
depend on the adopted reference abundances.
\begin{table}[h]
\begin{center}
\begin{tabular}{lcccc}\hline
Elem.&Adopted&$A_X$&$B_X$&$z_X$\\
$X$&$(X/{\rm H})_{\rm ref}$\\
\hline
  C&$8.46\pm 0.04$&$        -0.101\pm  0.229$&$        -0.193\pm  0.060$&  0.803\\
  N&$7.90\pm 0.11$&$        -0.000\pm  0.079$&$        -0.109\pm  0.111$&  0.550\\
  O&$8.76\pm 0.05$&$        -0.225\pm  0.053$&$        -0.145\pm  0.051$&  0.598\\
 Mg&$7.62\pm 0.02$&$        -0.997\pm  0.039$&$        -0.800\pm  0.022$&  0.531\\
 Si&$7.61\pm 0.02$&$        -1.136\pm  0.062$&$        -0.570\pm  0.029$&  0.305\\
  P&$5.54\pm 0.04$&$        -0.945\pm  0.051$&$        -0.166\pm  0.042$&  0.488\\
 Cl&$5.33\pm 0.06$&$        -1.242\pm  0.129$&$        -0.314\pm  0.065$&  0.609\\
 Ti&$5.00\pm 0.03$&$        -2.048\pm  0.062$&$        -1.957\pm  0.033$&  0.430\\
 Cr&$5.72\pm 0.05$&$        -1.447\pm  0.064$&$        -1.508\pm  0.055$&  0.470\\
 Mn&$5.58\pm 0.03$&$        -0.857\pm  0.041$&$        -1.354\pm  0.032$&  0.520\\
 Fe&$7.54\pm 0.03$&$        -1.285\pm  0.044$&$        -1.513\pm  0.033$&  0.437\\
 Ni&$6.29\pm 0.03$&$        -1.490\pm  0.062$&$        -1.829\pm  0.035$&  0.599\\
 Cu&$4.34\pm 0.06$&$        -0.710\pm  0.088$&$        -1.102\pm  0.063$&  0.711\\
 Zn&$4.70\pm 0.04$&$        -0.610\pm  0.066$&$        -0.279\pm  0.045$&  0.555\\
 Ge&$3.70\pm 0.05$&$        -0.615\pm  0.083$&$        -0.725\pm  0.054$&  0.690\\
 Kr&$3.36\pm 0.08$&$        -0.166\pm  0.103$&$        -0.332\pm  0.083$&  0.684\\
\hline
\end{tabular}
\caption{The depletion parameters for 17 elements.}
\label{parameters}
\end{center}
\end{table}

\vspace{-1.5cm}
\section{The Buildup of Dust Grains}\label{buildup}

\begin{figure}[hb]
\begin{center} 
\setlength\fboxsep{0pt}
\setlength\fboxrule{1.5pt}
\fbox{\includegraphics[width=5in]{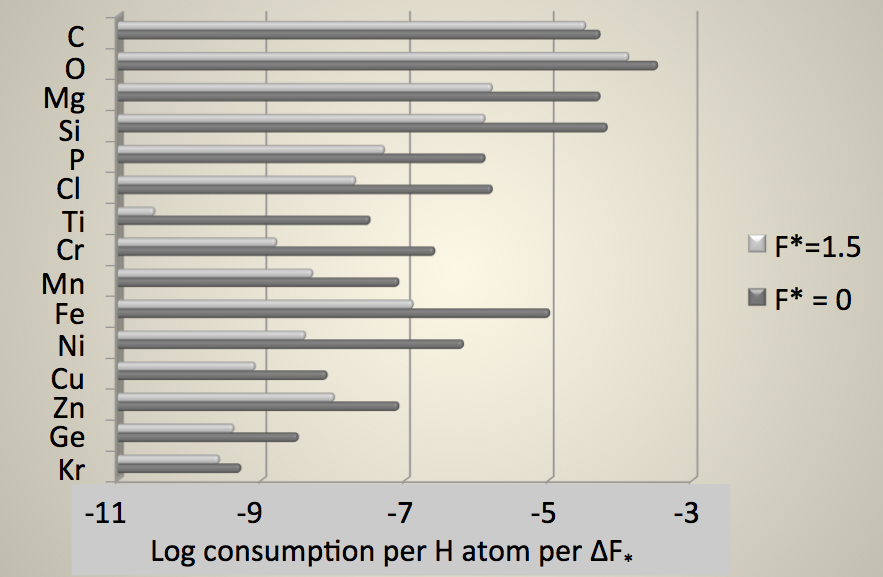}}
\caption{Logarithms of the element consumption rates $d(X_{\rm dust}/{\rm 
H})/dF_*$, as defined by Eq.~\protect\ref{differential_grain_comp}, for two values 
of $F_*$.  The elements here are the same as those listed in 
Table~\protect\ref{parameters}, except for nitrogen, which exhibits no 
measurable changes in depletion as a function of $F_*$ (i.e., 
$A_N=0.00$).\label{element_consumption}}
\end{center}
\end{figure}

A method for overcoming the effects arising from uncertainties in the reference 
abundances is to develop an interpretation that avoids having to use them in the 
first place.  Instead of attempting to derive the absolute quantities of elements 
sequestered in the grains, we can change our objective to that of monitoring the 
changes in element abundances as we progress from regions with low values of 
$F_*$ to ones where the grains have advanced to much greater stages of 
growth, as indicated by larger values of $F_*$.  If we substitute the right-hand 
side of the equation depicted in Fig.~\ref{basic_equation} into Eq.~\ref{x_dust} 
and differentiate the result with respect to $F_*$, we find that
\begin{eqnarray}\label{differential_grain_comp}
d(X_{\rm dust}/{\rm H})/dF_*&=&-(\ln 10)(X/{\rm H})_\odot 
A_X10^{B_X+A_X(F_*-z_X)}\nonumber\\
&=&-(\ln 10)A_X(X_{\rm gas}/{\rm H})_{F_*}
\end{eqnarray}
We now have an expression that makes use of only the slope $A_X$ and the 
(linear) expression for the relative abundance of the element $X$ with respect to 
hydrogen in the gas for any particular value of $F_*$.  The resulting consumption 
rates for different elements are shown in Fig.~\ref{element_consumption}.  These 
rates change as the grains mature, indicating that the compositions are driven by 
the abundance of the remaining gaseous feedstocks (free atoms) and perhaps 
also by the nature of the existing compounds in the grains with which new atoms 
can combine.  For instance, we can see from the differences in the bar lengths in 
the figure that there is little change in the consumption rates of C and O, 
presumably because they remain abundant even after many grains have formed, 
whereas a less abundant element that starts out being rapidly depleted, such as 
Ti, arrives at the point where there are few atoms left to contribute to further grain 
growth.  From the evidence that is presented here, it should be clear that the 
compositions of the outer mantles of grains differ from their inner cores that 
formed from gases that had higher concentrations of heavy atoms.
\section{Properties of the Sight Lines}\label{sight_lines}
\begin{figure}[h]
\includegraphics[scale=.5]{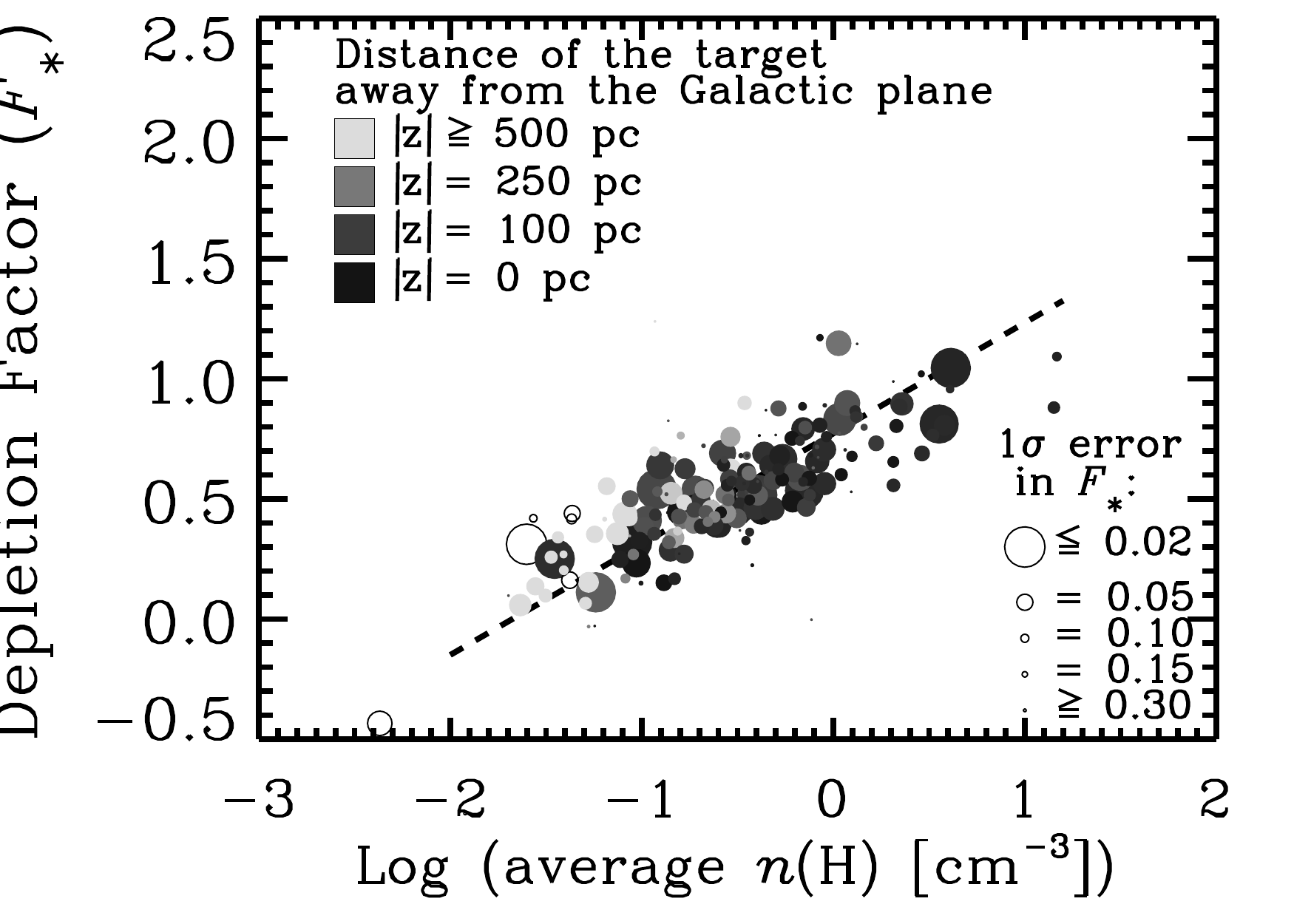}
\includegraphics[scale=.5]{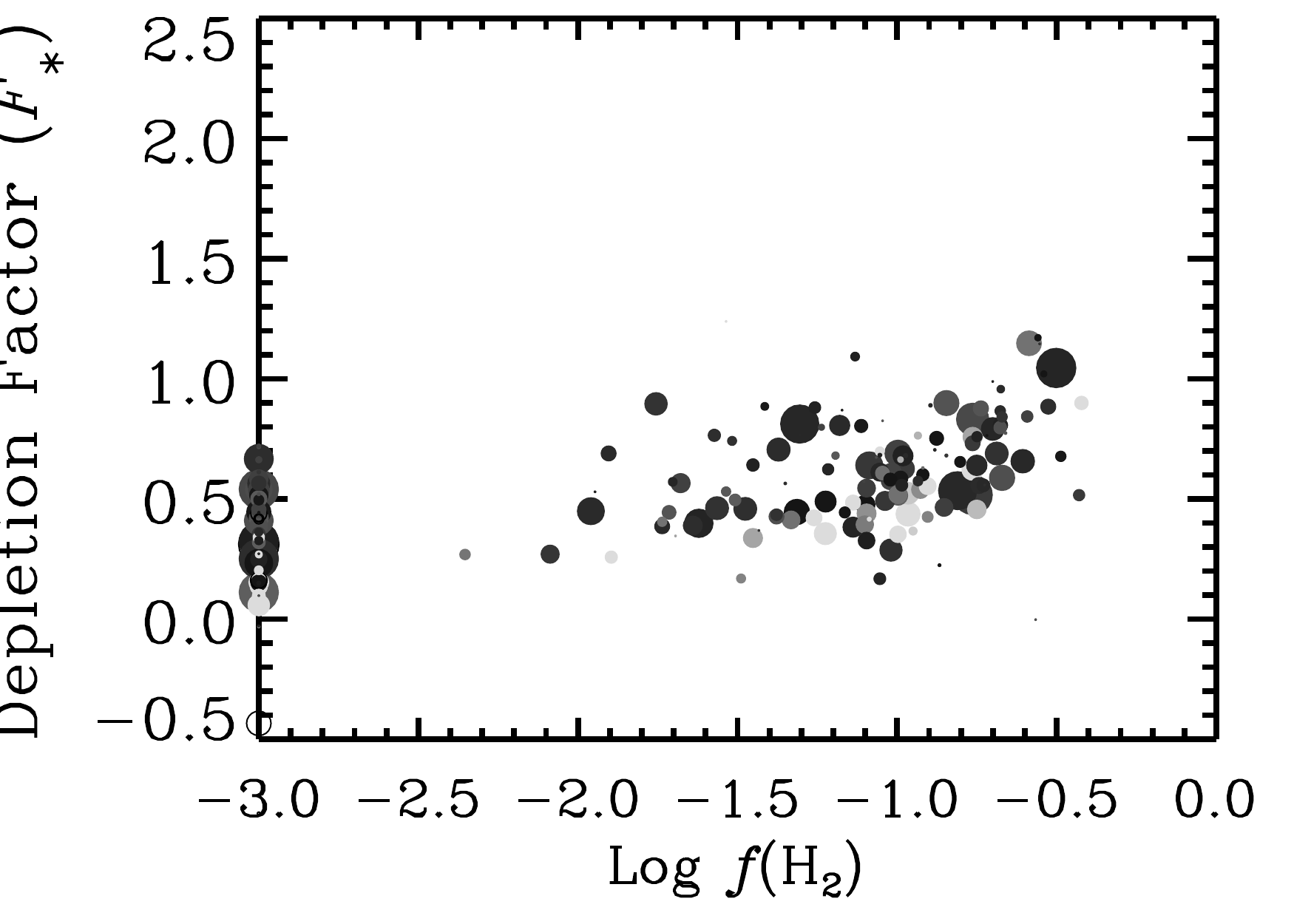}
\caption{{\it Left-hand panel:\/} The trend of $F_*$ as a function of the logarithm 
of the average density along the respective sight lines $\langle n({\rm 
H})\rangle$. As indicated by the legends, the sizes of the circles indicate the 
errors in $F_*$, and their shades of gray indicate the distances of the stars from 
the Galactic plane. Open circles indicate cases where $\log N({\rm H})<19.5$, 
where ionization effects may play a role in distorting the results.
The least-squares best fit to the trend is indicated by the dashed line.
{\it Right-hand panel:\/} The trend of $F_*$ against the logarithm of the fraction of 
hydrogen in molecular form $f({\rm H}_2)$. The sizes and gray levels of 
the points are the same as in the left-hand panel. Cases where $\log f({\rm 
H}_2)< -3.0$ are all grouped together on the $y$ axis of the 
plot.\label{trends}}
\end{figure}

In \S\ref{unified} we considered that the basic premise for the development of the 
$F_*$ parameter was that the collective depletion of many elements gives the 
most reliable indication of the overall level of grain production for any given sight 
line, after one applies proper scalings to account for the elements' depletion 
behaviors.  The small scatter in the relationships between the depletions of most 
elements and $F_*$ justifies this choice, which appears to be more satisfactory 
the more traditional methods of using the environmental factors $\langle n({\rm 
H})\rangle$ or $f({\rm H}_2)$ as guides for grain development.  Nevertheless, it is 
of interest to go back and see how well those measures compare with 
determinations of $F_*$.  Figure~\ref{trends} shows these relationships.  It 
appears that $\langle n({\rm H})\rangle$ is a better predictor for $F_*$ than 
$f({\rm H}_2)$.  This figure also shows that distances $|z|$ above or below the 
Galactic plane seem to have no systematic effects on the results.

\section{Comparisons with Condensation Temperatures}\label{Tc}

Another issue that we can revisit is how well do the element depletion factors 
scale with the condensation temperature $T_c$.  This quantity has a specific 
meaning related to the formation of solid compounds when a cosmic mixture of 
free atoms experiences a progressive decrease in temperature at some specific 
pressure, as in the expanding stellar atmospheres or nebulae considered by 
Field (1974).  There are, however, alternative processes that can be important.  
For instance, we can recognize that $T_c$ correlates well with other properties of 
atoms, such as ionization potentials (Snow 1975), binding energies to PAHs 
(Klotz et al. 1995), or sublimation energies (Barlow 1978) that govern the 
resistance of compounds to sputtering that can ultimately lead to the destruction 
of grain materials and the return of atoms to the gas phase.  Over any large 
selection of elements, these properties are correlated with each other (e.g. see 
Fig.~5 of Ramírez et al. (2010) for a plot of $T_c$ vs. first ionization potentials), 
so identifying the most important processes is not straightforward.  Nevertheless, 
we can propose that $T_c$ is a useful parameter to examine, even if it does not 
represent faithfully the most important formation and destruction routes.

\begin{figure}[h]
\hspace{-1cm}
 \includegraphics[scale=.25]{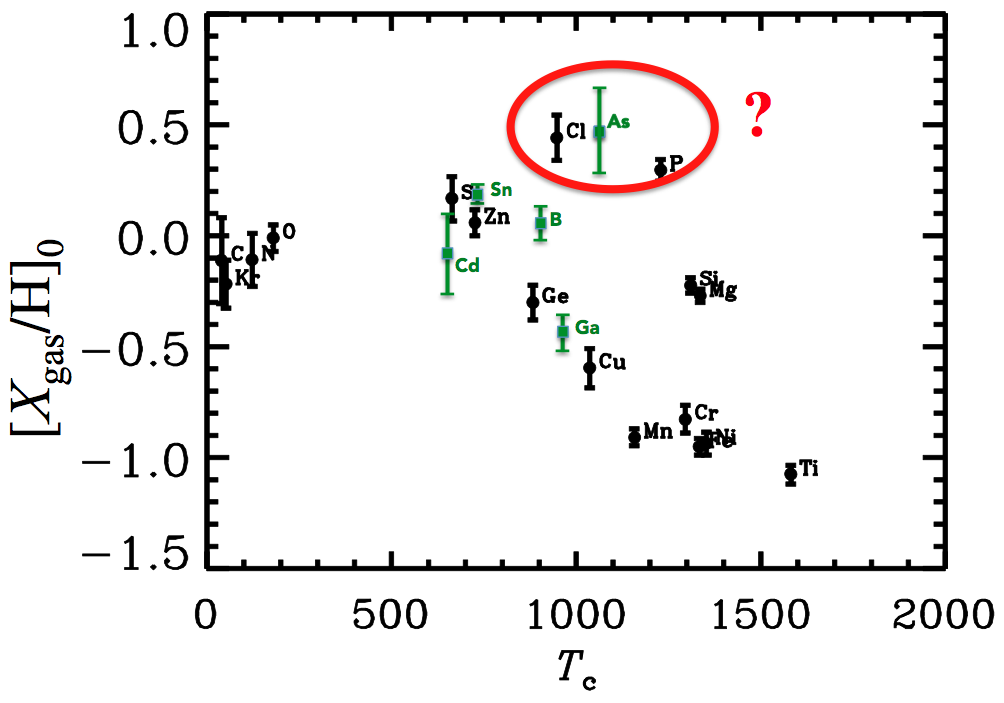}
\includegraphics[scale=.25]{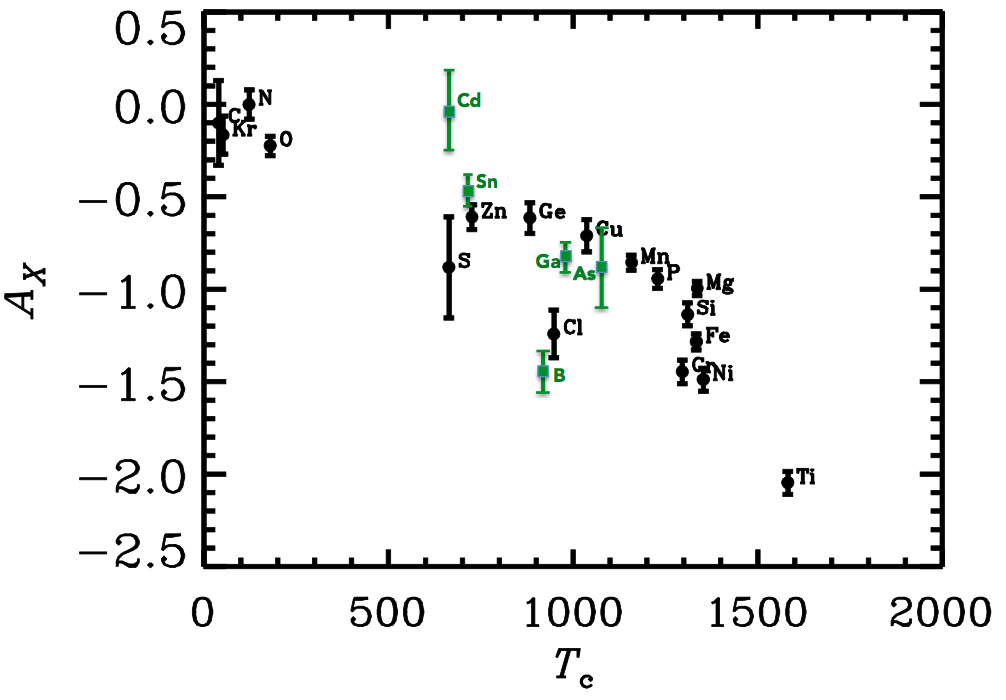}
\caption{{\it Left-hand panel:\/} The trend of element depletions for sight lines 
having $F_*\approx 0$ as a function of their respective condensation 
temperatures $T_c$. The 3 elements within the red ellipse appear to exhibit 
interstellar abundances that are {\it greater\/} than the reference abundances. {\it 
Right-hand panel:\/} The dependence of $A_X$, the slope of logarithmic 
depletions vs. $F_*$, on $T_c$.  In both panels, the elements shown in green are 
from a study of depletions by A.~M.~Ritchey (private 
communication).\label{Tc_trends}}
\end{figure}

The left-hand panel of Figure~\ref{Tc_trends} shows how well the depletions at 
$F_*=0$ correlate with $T_c$.  While a trend seems evident, there are some 
awkward cases that involve 3 elements where {\it negative\/} depletions are 
apparent (i.e., abundances that seem to be greater than their respective 
reference abundances).  An explanation for these anomalies may rest with either 
bad measurements, incorrect transition $f$-values, or inaccurate reference 
abundances.  A more satisfactory trend is seen when $A_X$ (the slope of 
logarithmic depletions vs. $F_*$) is compared with $T_c$.  Here, the 
dependences on $f$-values or the adopted reference abundances are absent.

Recent studies that compare the differences of abundances in the atmosphere of 
the Sun compared to those of solar twins and solar analogs(Meléndez et al. 2009 
; Ramírez et al. 2009)\footnote{Ramírez, et al. (2009) define a solar twin as a star 
that has $T_{\rm eff}$ within 100$\,$K, $\log g$ within 0.1$\,$dex, and [Fe/H] 
within 0.1$\,$dex of the respective values of the Sun.  Solar analogs are main-
sequence stars that have spectral types G0$-$G5 (Meléndez, et al. 2009).} show 
a pattern very similar to the $A_X$ vs. $T_c$ trend shown in 
Fig.~\ref{Tc_trends}, but much smaller in amplitude.  A popular conjecture is that 
the refractory elements condense out to form dusty material that eventually goes 
into rocky planets or the interiors of gas giants, and that the remaining gas is 
dumped into a thin convection zone of the star.  However, Ramírez et al. (2014) 
surveyed the differential abundances of many solar-type stars and were not able 
to find a good correlation of these depletions with the presence or absence of 
planets; the picture may be muddled by how deep the stars' convection zones 
were at the time when the separations took place and the planetary disks formed.

\section{Comments about Specific Elements}\label{elements}
A number of interesting conclusions emerge from the findings for specific 
elements, as summarized in the subsections that follow.

\subsection{Carbon}\label{C}

Many investigators (Snow \& Witt 1995 ; Cardelli et al. 1996 ; Kim \& Martin 1996 
; Mathis 1996 ; Dwek 1997 ; Li 2005) have expressed concerns that the C 
depletion results indicated that there was an insufficient amount of carbon in solid 
form to reconcile the observed optical and near-UV extinctions with calculations 
of the optical properties of dust grains, where the requirements for the fractional 
amounts of C in solids ranged from $150-200$ parts per million H atoms (ppm) 
(Mathis 1996) to 300$\,$ppm (Draine \& Lee 1984). This has often been referred 
to as the ``carbon crisis.''  If we apply Eq.~\ref{x_dust} to the parameters shown 
in Table~\ref{parameters}, we find that the amount of C locked up in dust grains 
ranges from $({\rm C}_{\rm dust}/{\rm H})=65\,$ppm for $F_*=0.0$ to 111$\,$ppm 
for $F_*=1.0$.  From this, one might conclude that the carbon crisis still remains.  
However, Sofia et al. (2011) have measured the damping wings of the 
1334$\,$\AA\ allowed transition of C~II in some spectra instead of the 
customarily-measured strengths of the very weak intersystem line at 
2325$\,$\AA.  They obtained interstellar carbon column densities that were about 
0.43 times as large as those obtained from the weak line.  If the revision 
suggested by Sofia, et al. is correct, then the carbon crisis may be mostly 
resolved, since now $({\rm C}_{\rm dust}/{\rm H})=(192,\, 212)\,$ppm for the two 
values of $F_*$.  It is generally true that the $f$-values that are calculated for 
semi-forbidden lines are much less reliable than those for allowed transitions, 
which may indicate that the new C abundances are more trustworthy.  However, 
an opposing consideration is the fact that Sofia, et al. had to measure damping 
wings on top of strong stellar C~II lines, which they had to estimate from 
calculations of stellar model atmospheres and radiative transfer.

\subsection{Nitrogen}\label{N}

Of all 17 elements studied by Jenkins (2009), nitrogen appears to be only one 
that shows no measurable change in depletion as $F_*$ is increased.  The 
entries for nitrogen in Table~\ref{parameters} indicates that $[{\rm N}_{\rm 
gas}/{\rm H}]$ appears to hold at a steady value of $-0.109\pm0.111\,$dex, which 
is possibly consistent with no depletion at all.  Gail and Sedlmayr (1986) 
suggested that the condensation of N into any sort of solid compound might be 
inhibited by the production of ${\rm N}_2$ which has a saturated bond with a very 
high activation energy for any gas-phase reactions that could form other 
molecules.  This may be an attractive theoretical consideration, but the observed 
abundance of N$_2$ in the interstellar medium is small (Knauth et al. 2004, 
2006).

\subsection{Oxygen}\label{O}

A conventional view is that oxygen in dust grains appears in the form of refractory 
compounds such as silicates and metallic oxides.  In this category, the elements 
Mg, Si, and Fe are the only ones with significantly high abundances to be 
relevant.  We can examine whether or not the \underbar{\emph{differential}} 
depletions O with respect to these three elements is consistent with this picture 
(again, this allows us to sidestep the uncertainties in the reference abundance of 
O, which, through the years, has shown large changes from one determination to 
the next;  see \S3.4 of ref (Asplund, et al. 2009) for details).  At $F_*=0.0$, 
Eq.~\ref{differential_grain_comp} and the parameters in Table~\ref{parameters} 
tell us that $d({\rm O}_{\rm dust}/{\rm H})/dF_*=1.6d({\rm Mg+Si+Fe}_{\rm 
dust}/{\rm H})/dF_*$, which is consistent with the most oxygen-rich silicate, 
MgSiO$_3$ where the ratio of O to the other atoms is 1.5.  However, when 
$F_*=1.0$ the differential consumption of O relative to the other three elements is 
ten times higher!  Even when we consider the $-2\sigma$ uncertainty in $d({\rm 
O}_{\rm dust}/{\rm H})/dF_*$, we find that the ratio only drops to 6 times the 
consumption rate of Mg + Si + Fe.

Clearly, in addition to forming silicates or metallic oxides, oxygen must be in 
bound form with some other \underbar{\emph{abundant}} element.  We can rule 
out oxides of nitrogen, because the differential depletion of this element is about 
zero (see \S\ref{N}).  Water ice seems to be an attractive prospect; it is known to 
be abundant in very dense clouds and protostellar nebulae (van Dishoeck 2004), 
but various surveys of the 3.6$\mu$m ice band shows a linear trend with visual 
extinction $A_V$ that extrapolates to zero for $A_V=2.6-5$ (Whittet et al. 1988 ; 
Eiroa \& Hodapp 1989 ; Smith et al. 1993) (most $A_V$ values for the sight lines 
in the UV surveys are less than 1).  The molecules CO, CO$_2$ and O$_2$ are 
present in the interstellar medium, but not in amounts that are significant enough 
to explain the depletion of oxygen.  Whittet (2010) has examined the oxygen 
depletion problem in some detail and concludes that organic refractory 
compounds may explain the extraordinary consumption of oxygen.  Another 
possible solution might be that outside of the dense clouds H$_2$O ice may still 
be retained on grains of large diameter ($>1\mu$m) that are opaque to infrared 
radiation and hence would not exhibit the 3.6$\mu$m absorption feature.

\begin{figure}[h]
\begin{center}
\includegraphics[scale=.25]{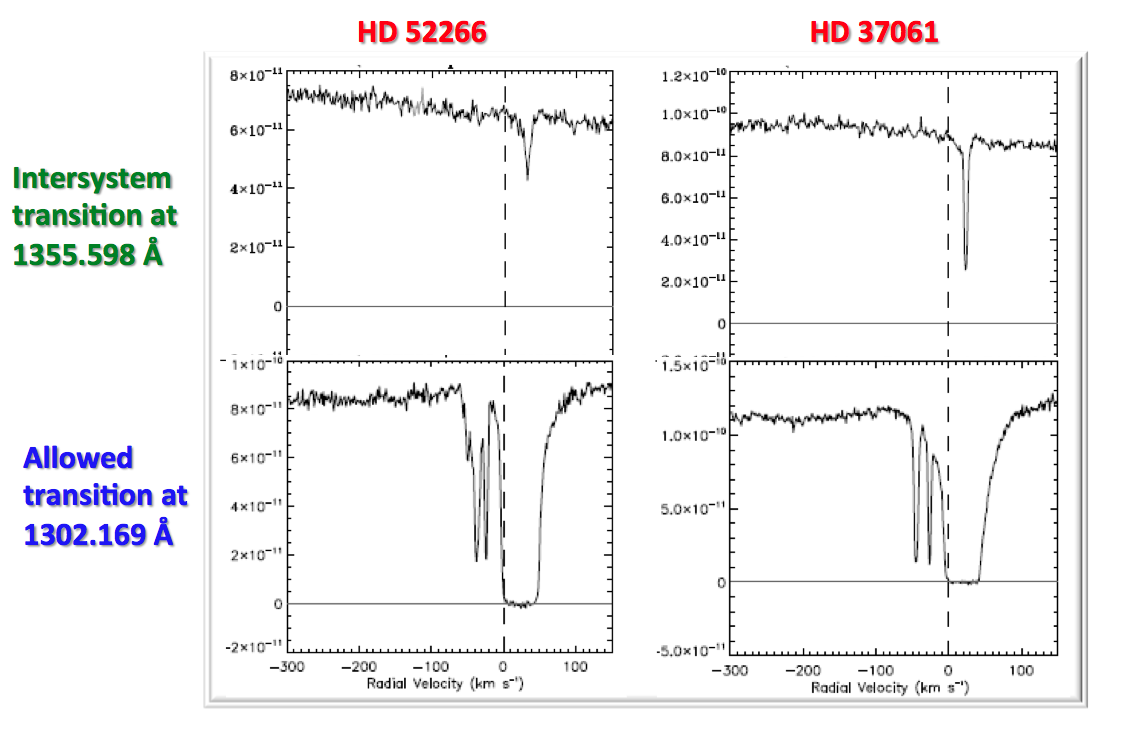}
\caption{Recordings of spectral segments made by STIS on HST for two stars 
(labeled at the top).  The strengths of the absorptions by the very weak transition 
at 1356$\,$\AA\ ({\it top panels\/}) are in agreement with the damping wings of 
the very strongly saturated line at 1302$\,$\AA\ ({\it bottom 
panels\/}).\label{oi_lines}}
\end{center}
\end{figure}

After having read \S\ref{C}, one might question whether or not the determinations 
of the abundance of O in the gas phase could be wrong because the adopted 
$f$-value\footnote{Jenkins (2009) corrected all of the O~I determinations based 
on the 1356$\,$\AA\ line so that they were consistent with an $f$-value equal to 
$1.16\times 10^{-6}$ given in the compilation of Morton (2003).} of the weak line 
at 1356$\,$\AA\ was in error, much like the indications that we had with the 
2325$\,$\AA\ line of C~II.  A large fraction of the oxygen measurements were 
made with this intersystem line.  Figure~\ref{oi_lines} shows both the 
1356$\,$\AA\ absorption and that from the permitted line at 1302$\,$\AA\ for two 
sight lines where high velocity features do not compromise our ability to see the 
damping wings of the strong line.  A preliminary comparison of the two by the 
author indicates that the outcomes for the two lines show exceptionally good 
agreement with each other.

An entirely different way to sense the amount of oxygen in bound form is to 
examine the detailed structure of the K-edge X-ray absorption feature at a 
wavelength of 23$\,$\AA.  When the oxygen atoms are chemically bound to other 
atoms, the energies of their electron states are perturbed slightly.  The analysis 
of the absorption structures in the vicinity of the K-edge is rather complicated, 
owing to the presence of many additional features from higher ionization levels of 
free oxygen atoms (Juett et al. 2004).  Pinto et al. (2013) examined the X-ray 
spectra of nine Galactic low-mass X-ray binaries obtained from the Reflection 
Grating Spectrometer on {\it XMM-Newton}.  They found that the percentage of 
oxygen in chemically bound form ranges between 15 and 25\% of the total 
amount of oxygen present.  However, degeneracies in the signatures of different 
compounds prevented them from identifying the detailed chemical structure of 
oxygen-bearing substances.  Their reported range for the percentage of bound 
oxygen corresponds to the UV absorption line results for $0.26<F_*<0.51$.

\subsection{Krypton}\label{Kr}

A surprising result to emerge from the study of krypton abundances is that $[{\rm 
Kr_{gas}/H}]$ varies from an extreme of $-0.5$ to $-0.1$ (see the right-most 
panel in Fig.~\ref{basic_equation}).  In the latest of a series of investigations of 
Kr, Cartledge et al. (2008) dismissed the notion that Kr could be depleted onto 
grains because it is a noble gas.  Instead, they viewed the variations in Kr 
abundances as a consequence of differing contributions from nucleosynthetic 
activity in different regions of space.  While this could remain a possibility, it is 
nevertheless evident that there is a trend that relates $[{\rm Kr_{gas}/H}]$ to 
$F_*$, but only at an 89\% level of confidence based on a Pearson correlation 
coefficient (with a one-tail test because we can reject positive values of $A_{\rm 
Kr}$; also note from Table~\ref{parameters} that while $A_{\rm Kr}$ is negative 
at only the $1.6\sigma$ level of significance,\footnote{A. M. Ritchey (private 
communication) has remeasured many Kr abundances, and his new results 
strengthen the significance of the negative slope for $A_{\rm Kr}$.} $B_{\rm Kr}$ 
is negative at the $4\sigma$ level).  No trend was evident when Cartledge, et al. 
(2008) plotted $[{\rm Kr_{gas}/H}]$ against $\langle n_{\rm H}\rangle$, which 
indicates that the average density along a sight line is probably a less accurate 
indicator of depletion than $F_*$, which relies on elements with highly significant 
depletions as indicators.

It may be possible to understand the krypton depletions in terms of it binding 
either in the form of a clathrate hydrate (Lodders 2003) or in a complex with ${\rm 
H}_3^+$ (Mousis et al. 2008).  For the former, recall that in \S\ref{O} we 
considered the possibility that large quantities of water ices may be on the 
surfaces of grains with large diameters.  For the latter, we know that cosmic ray 
ionization can result in the creation of fractional abundances of ${\rm H}_3^+/{\rm 
H}_2$ as high as $10^{-7}$ in low density regions of the interstellar medium 
(Indriolo \& McCall 2012).  For either of these means for sequestering Kr, it would 
be useful to perform calculations of the equilibrium concentrations that might be 
expected.

\section{Concluding Remarks}\label{concluding_remarks}

A study of the depletions of different elements from the gas phase of the 
interstellar medium sets some important constraints on the possible compositions 
of dust grains in different stages of development or levels of concentration.  
However, we do not obtain explicit information on what compounds are formed.  
One can, however, make some illustrative guesses on the possible mixes of 
minerals that satisfy the constraints, as had been done by Draine (2004), who 
examined the results from a contemporaneous study of depletions by Jenkins 
(2004).

Some words of caution are in order.  There may be some dangers in
over-generalizing the results presented in \S\S \ref{unified}$-$\ref{buildup} when 
one considers grain compositions in systems other than our immediate 
neighborhood in our Galaxy.  One might suppose that in a system where the 
proportions of different elements are not the same as what we see here, it would 
be a simple matter to estimate how rapidly the elements disappear into solid form 
by just integrating the differential depletion relationships, as expressed in 
Eq.~\ref{differential_grain_comp}, after starting with some initial gas composition.  
Complications arise when we recognize that some compounds rely on the 
presence of others to form initially and remain stable for long periods of time. For 
instance, Lodders (2003) has pointed out that the incorporation of the elements 
Ni and Ge in solids depends on the presence of a host element Fe to create an 
alloy.  Likewise, the formation of refractory compounds that contain Zn and Mn, 
such as ${\rm Zn_2SiO_4}$, ${\rm ZnSiO_3}$, or ${\rm Mn_2SiO_4}$, are aided 
by pre-existing host minerals such as forsterite and enstatite.  In some galaxy 
environment where the $\alpha/{\rm Fe}$ differs from that of our own, some 
elements may have more or less than their respective host compounds that 
contain other key elements, thus altering the depletion rates in a way that is 
difficult to predict.  We need also to recognize that any changes in the mix of 
sources that create the initial grains, such as different kinds of stars and 
supernovae, can have a profound influence on the outcome.  These initial grains 
are important constituents by themselves, and moreover they eventually form the 
cores of grains that undergo further development in the interstellar medium.

\end{document}